\documentclass[aps,prb,twocolumn,showpacs,superscriptaddress]{revtex4}
\usepackage{amsfonts}
\usepackage{amsmath}
\usepackage{amssymb}
\usepackage{graphicx}
\usepackage{float}

\begin{document}

\title{Efficient spin-current injection in single-molecule magnet junctions}
\author{Haiqing Xie}
\affiliation{Department of Physics, Taiyuan Normal University, Taiyuan 030001, China}
\author{Fuming Xu}
\email{xufuming@szu.edu.cn}
\affiliation{College of Physics and Energy, Shenzhen University, Shenzhen 518060, China}
\author{Hujun Jiao}
\affiliation{Institute of Theoretical Physics, Shanxi University, Taiyuan 030006, China}
\author{Qiang Wang}
\email{q2wang332@163.com}
\affiliation{Department of Physics, Taiyuan Normal University, Taiyuan 030001, China}
\author{J.-Q. Liang}
\affiliation{Institute of Theoretical Physics, Shanxi University, Taiyuan 030006, China}

\begin{abstract}
We study theoretically spin transport through a single-molecule magnet (SMM)
in the sequential and cotunneling regimes, where the SMM is weakly coupled
to one ferromagnetic and one normal-metallic leads. By a master-equation
approach, it is found that the spin polarization injected from the
ferromagnetic lead is amplified and highly polarized spin-current can be
generated, due to the exchange coupling between the transport electron and
the anisotropic spin of the SMM. Moreover, the spin-current polarization can
be tuned by the gate or bias voltage, and thus an efficient spin injection
device based on the SMM is proposed in molecular spintronics.
\end{abstract}

\date{\today }
\maketitle

\section{Introduction}

As is well known, the efficient generation and manipulation of
spin-polarized current are the crucial elements in the field of spintronics.
\cite{spin} Considerable efforts have been devoted to direct spin-current
injection from ferromagnetic (FM) electrodes into nonmagnetic materials via
tunnel junctions, such as graphene,\cite{graphene1,graphene2} silicon,\cite%
{silicon} and quantum dots (QDs).\cite{QDE1,QDE2,QDT1,QDT2} It is shown both
experimentally\cite{QDE1,QDE2} and theoretically\cite{QDT1} that the
spin-current polarization can be electrically manipulated in transport
through a single QD weakly coupled to one FM and one normal-metallic (NM)
electrodes, but the polarization is limited by the typical polarization of
ferromagnet with $30-40\%$.\cite{sp} Interestingly, when the coupling of the
FM lead with the QD is much stronger than that of the NM lead, the
polarization of spin injection in the strong coupling regime is greatly
enhanced, far beyond the intrinsic polarization of ferromagnets,\cite{QDT2}
due to the FM exchange-field induced spin-splitting of the QD level.
However, high spin polarization strongly depends on the left-right asymmetry
of the QD-lead coupling in this case.

In molecular spintronics, single-molecule magnets (SMMs) with a large
molecular spin have magnetic bistability induced by the easy-axis magnetic
anisotropy, which has potential application in data storage and information
processing. Therefore, transport properties through SMMs have been
intensively investigated in recent years.\cite{Wernsdorfer,exp1,Kim} For
instance, the spin-filter effect\cite{Spinfilter1,Spinfilter2} and
thermoelectrically induced pure spin-current\cite{Xing2,Xing3} are
identified in SMM junctions with NM leads. When the SMM is attached to two
FM electrodes, the tunnel magnetoresistance,\cite{JBTMR1,Xie1,JBTMR2,JBTMR3}
memristive properties,\cite{Timm3} spin Seebeck effect,\cite%
{SSeebeck,SSeebeck1} spin-resolved dynamical conductance,\cite{Scon} and
other spin-related properties have been theoretically studied. Due to the
spin asymmetry, the spin-polarized transport through a SMM coupled to one FM
and one NM leads has received much attention. It is both theoretically\cite%
{Timm1,JB,Rossier,Xing1,JB1} and experimentally\cite{exp2,exp3} verified
that, the spin switching of such FM-SMM-NM junction can be realized by
spin-polarized currents. On the other hand, the spin-polarized charge
current itself can exhibit behaviors such as the spin-diode\cite{JBdiode}
and negative differential conductance.\cite{Timm2,Xue} However, the
spin-current through the magnetic molecular junction is less mentioned in
these works.

In this paper, we adopt the master-equation approach to study spin-current
injection through the FM-SMM-NM junction in the weak coupling regime. The
system under investigation is shown in Fig.1, where the magnetization of the
FM lead is collinear with the magnetic easy axis of the SMM. Experimentally,
this transport setup can be realized on a FM scanning tunneling microscope
tip coupled to a magnetic adatom or SMM placed on a NM surface.\cite%
{exp1,exp2,Rossier,JBdiode,Xie2} The SMM is modeled as a single-level QD
with a local uniaxial anisotropic spin.\cite{Timm1,JBTMR1} Both FM and
antiferromagnetic (AFM) exchange couplings are discussed in this work. We
find that the output spin-current polarization through the SMM junction is
greatly enhanced and can reach $90\%$, where spin polarization is typically $%
40\%$ in the FM lead. The enhancement of spin polarization is attributed to
the easy-axis magnetic anisotropy of the SMM and spin-flip process.\cite%
{JBTMR1,Xie1,Xie2,Xie3}

\begin{figure}[tph]
\begin{center}
\includegraphics[width=0.9\columnwidth]{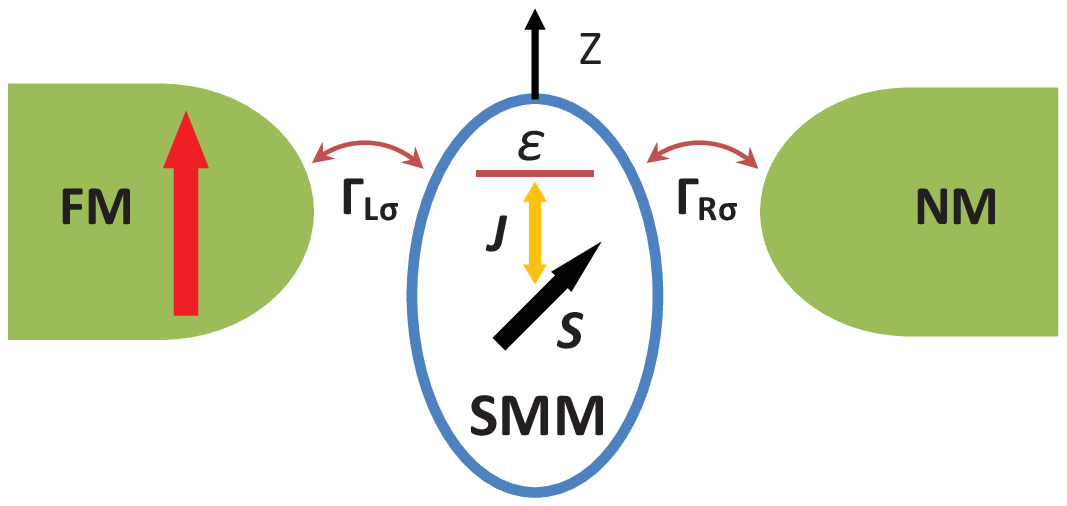}
\end{center}
\caption{(Color online) Schematic diagram for a SMM weakly coupled to FM and
NM electrodes. The magnetization of FM lead is collinear with the easy axis
of SMM (assumed as $z$-axis). Bias voltages $V/2$ and $-V/2$ are applied on
the left (L) and right (R) electrodes, respectively.}
\label{Fig:1}
\end{figure}

\section{Model and method}

The total Hamiltonian of the SMM tunnel junction shown in Fig. 1 is written
as%
\begin{equation}
H=H_{SMM}+H_{leads}+H_{T}.  \label{Eq1}
\end{equation}%
The first term is the giant-spin Hamiltonian of the SMM\cite%
{Timm1,JBTMR1,Shen}%
\begin{eqnarray}
H_{SMM} &=&\sum_{\sigma }(\varepsilon -eV_{g})d_{\sigma }^{\dag }d_{\sigma
}+Ud_{\uparrow }^{\dag }d_{\uparrow }d_{\downarrow }^{\dag }d_{\downarrow }
\notag \\
&&-K(S^{z})^{2}-J\mathbf{s\cdot S}.
\end{eqnarray}%
Here, $\varepsilon $ is the energy of the orbital level (OL) of the magnetic
molecule, which can be tuned by the gate voltage $V_{g}$, and the operator $%
d_{\sigma }^{\dag }$ ($d_{\sigma }$) creates (annihilates) an electron with
spin $\sigma $ in the molecular OL. $U$ is the Coulomb energy of the two
electrons in the molecule, and $K$ ($K>0$) denotes the easy-axis anisotropy
of the SMM. The spin operator of the OL is defined as\textbf{\ }$\mathbf{%
s\equiv }$ $\sum_{\sigma \sigma ^{\prime }}d_{\sigma }^{\dag }(\mathbf{%
\sigma }_{\sigma \sigma ^{\prime }}/2)d_{\sigma ^{\prime }}$, where $\mathbf{%
\sigma \equiv (\sigma }_{x},\mathbf{\sigma }_{y},\mathbf{\sigma }_{z}\mathbf{%
)}$ represents the vector of Pauli matrices. $J$ describes the spin-exchange
coupling between spin $\mathbf{s}$ of OL electrons and the local spin $%
\mathbf{S}$ of the molecule, which can be either FM ($J>0$) or AFM ($J<0$).
By introducing the $z$ component $S_{t}^{z}$ of the total spin operator $%
\mathbf{S}_{t}\mathbf{\equiv s+S}$, many-body eigenstates of the SMM can be
written as $\left\vert n,S_{t};S_{t}^{z}\right\rangle $, where $n$ denotes
the charge state of the SMM and $S_{t}$ is the quantum number of the total
spin $\mathbf{S}_{t}$.

The second term of Eq.(\ref{Eq1}) describes noninteracting electrons in the
electrodes, $H_{leads}=\sum_{\mathbf{\alpha =L,R}}\sum_{\mathbf{k}\sigma
}\varepsilon _{\alpha \mathbf{k\sigma }}c_{\alpha \mathbf{k}\sigma }^{\dag
}c_{\alpha \mathbf{k}\sigma }$, where $\varepsilon _{\alpha \mathbf{k}\sigma
}$ is the energy of an electron with wave vector $\mathbf{k}$ and spin $%
\sigma $ in lead $\alpha $, and $c_{\alpha \mathbf{k}\sigma }^{\dag }$ ($%
c_{\alpha \mathbf{k}\sigma }$) is the corresponding electronic creation
(annihilation) operator. Assuming $\rho _{\alpha \sigma }$ is the density of
states of electrons with spin $\sigma $ in the lead $\alpha $, we can define
the spin polarization of the ferromagnetic lead as $p_{\alpha }=(\rho
_{\alpha \sigma }-\rho _{\alpha \overline{\sigma }})/(\rho _{\alpha \sigma
}+\rho _{\alpha \overline{\sigma }})$. In our calculation, polarizations of
the left FM and right NM leads are chosen as $p_{L}=0.4$ and $p_{R}=0$,
respectively.

The coupling between the leads and the SMM is described by the tunneling
Hamiltonian $H_{T}=\sum_{\alpha \mathbf{k}\sigma }(t_{\alpha }c_{\alpha
\mathbf{k}\sigma }^{\dag }d_{\sigma }+t_{\alpha }^{\ast }d_{\sigma }^{\dag
}c_{\alpha \mathbf{k}\sigma })$, where $t_{\alpha }$ is the tunnel matrix
element between the lead $\alpha $ and the SMM, and the spin-dependent
tunnel-coupling strength is denoted by $\Gamma _{\alpha \sigma }=2\pi \rho
_{\alpha \sigma }\left\vert t_{\alpha }\right\vert ^{2}$. Furthermore, we
can rewrite the tunnel-coupling strength as $\Gamma _{\alpha \sigma }=\Gamma
_{\alpha }(1\pm p_{\alpha })/2$ with the sign $+$ ($-$) corresponding to $%
\sigma =\uparrow $ ($\downarrow $), and assume $\Gamma _{\alpha }=\Gamma
_{\alpha \uparrow }+\Gamma _{\alpha \downarrow }$ and $\Gamma =(\Gamma
_{L}+\Gamma _{R})/2$. For simplicity, the bias voltage $V$ is applied
symmetrically on the SMM tunnel junction with $\mu _{L}=eV/2$ and $\mu
_{R}=-eV/2$.

\begin{figure}[pth]
\begin{center}
\includegraphics[width=1\columnwidth]{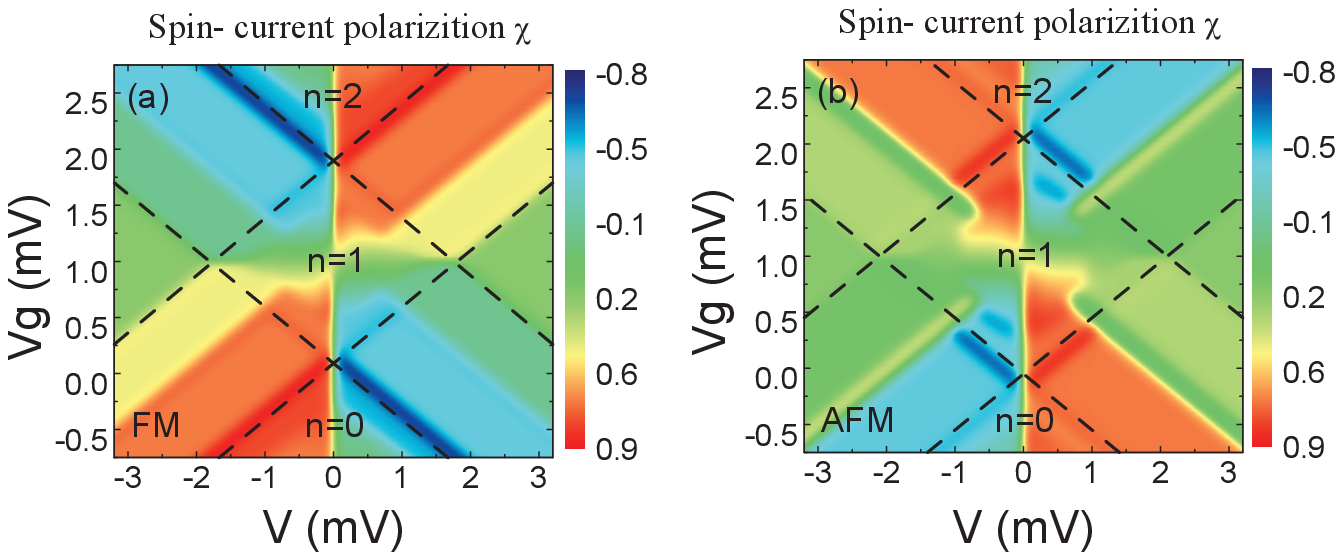}
\end{center}
\caption{(Color online) Spin-current polarization for the FM (a) and AFM (b)
exchange couplings as a function of the bias and gate voltages. The
parameters are: $S=2$, $\protect\varepsilon=0.5$ meV, $|J|=0.4$ meV, $U=1$
meV, $K=0.05$ meV, $k_{B}T=0.04$ meV, $P_{L}=0.4$, $P_{R}=0$, and $%
\Gamma=\Gamma_{L}=\Gamma_{R}=0.001$ meV.}
\label{Fig:2}
\end{figure}

We analyze spin-polarized transport through the SMM junction in both
sequential and cotunneling regimes at the weak-coupling limit, ie. $\Gamma
\ll k_{B}T$. The tunneling processes of electron are assumed to be
stochastic and Markovian, and the time evolution of the SMM can be described
by the master equation,\cite{Xie1,Xie3,Timm5,Koch2}%
\begin{eqnarray}
\frac{dP_{i}}{dt} &=&\sum_{\alpha \alpha ^{\prime }\sigma \sigma ^{\prime
}i^{\prime }\neq i}[-(W_{\alpha \sigma ,\alpha ^{\prime }\sigma ^{\prime
}}^{i,i^{\prime }}+W_{\alpha \sigma }^{i,i^{\prime }})P_{i}  \notag \\
&&+(W_{\alpha \sigma }^{i^{\prime },i}+W_{\alpha ^{\prime }\sigma ^{\prime
},\alpha \sigma }^{i^{\prime },i})P_{i^{\prime }}],  \label{master}
\end{eqnarray}%
with $P_{i}$ denoting the probability in the molecular many-body eigenstate $%
\left\vert i\right\rangle $. The transition rates $W$ in the Eq.~(\ref%
{master}) can be calculated perturbatively by the generalized Fermi's golden
rule based on the $T$-matrix.\cite{Bruus} Moreover, the rate $W_{\alpha
\sigma }^{i,i^{\prime }}$ denotes the sequential tunneling transition from
the state $\left\vert i\right\rangle $ to $\left\vert i^{\prime
}\right\rangle $ due to a spin-$\sigma $ electron tunneling of lead $\alpha $%
, and $W_{\alpha \sigma ,\alpha ^{\prime }\sigma ^{\prime }}^{i,i^{\prime }}$
stands for the cotunneling transition with a spin-$\sigma $ electron of lead
$\alpha $ being transformed to spin-$\sigma ^{\prime }$ electron of lead $%
\alpha ^{\prime }$. With the stationary conditions $\frac{%
dP_{i}}{dt}=0$ and $\sum_{i}P_{i}=1$, we can get the steady state transport
properties. Finally, the current of spin-$\sigma $ electrons through lead $%
\alpha $ is defined as\cite{Xie3,Koch2}%
\begin{align}
I_{\alpha \sigma }& =e(-1)^{\delta _{R\alpha }}\sum_{\alpha ^{\prime }\neq
\alpha \sigma ^{\prime }ii^{\prime }}[(n_{i^{\prime }}-n_{i})W_{\alpha
\sigma }^{i,i^{\prime }}P_{i}  \notag \\
& +(W_{\alpha \sigma ,\alpha ^{\prime }\sigma ^{\prime }}^{i,i^{\prime
}}-W_{\alpha ^{\prime }\sigma ^{\prime },\alpha \sigma }^{i,i^{\prime
}})P_{i}],
\end{align}%
and thus we have the total charge current $I_{\alpha }=I_{\alpha \uparrow
}+I_{\alpha \downarrow }$ as well as the spin-current $I_{\alpha
s}=I_{\alpha \uparrow }-I_{\alpha \downarrow }$. The polarization of
spin-current is defined as
\begin{equation*}
\chi =(I_{\alpha \uparrow }-I_{\alpha \downarrow })/(I_{\alpha \uparrow
}+I_{\alpha \downarrow })
\end{equation*}%
In addition, magnetization of the SMM is $\left\langle
S_{t}^{z}\right\rangle =\sum_{i}S_{ti}^{z}P_{i}$. With this well-defined
theory, we can calculate the spin transport properties of the SMM junction
and the results are presented below.

\section{Results and discussion}

In Fig. 2, we show the dependence of the spin-current polarization on the
gate and bias voltages for the FM and AFM spin-exchange couplings. We first
find that the polarization $\chi $ is asymmetric under reversal of the bias
voltage $V$, since tunnel-coupling strengths between the molecule and the FM
electrode is spin-dependent ($\Gamma _{L\uparrow }>\Gamma _{L\downarrow }$).
At lower bias voltages, the cotunneling processes dominates electron
transports in the regions marked by the electron occupation numbers $n=0$ ($%
1 $ or $2$),\cite{JBTMR1,Xie1} and the sequential tunneling processes start
to take part in transports at larger bias voltages. Although the
polarization of FM lead is 0.4, the spin-current polarization $\chi $ can be
beyond $0.4$, or even more than $0.9$ in the sequential and cotunneling
regions [Figs. 2(a) and (b)]. Moreover, because the energy of the molecular
state $\left\vert 1,5/2;S_{t}^{z}\right\rangle $ is larger than that of the
molecular state $\left\vert 1,3/2;S_{t}^{z}\right\rangle $ for the AFM case,%
\cite{Xie3} the voltage-dependence of spin-polarization $\chi $ is very
different for the two types of couplings, or even opposite in some regions.

The charge current $I_{c}$, spin current $I_{s}$, differential conductance $%
G $, magnetization $\left\langle S_{t}^{z}\right\rangle $, and spin
polarization $\chi $ are shown in Fig. 3 for the FM spin-exchange coupling.
The spin-flip induced by the spin-exchange coupling plays a key role \cite%
{JBTMR1,Xie1,JBdiode,Xie3} in transport through the SMM. The
magnetization $\left\langle S_{t}^{z}\right\rangle $ is a positive value in
the negative bias voltages [Fig. 3(c)], since the states with positive $%
S_{t}^{z}$ dominate the steady transport processes. It becomes negative in
the positive bias voltages. The charge current $I_{c}$ (black solid line)
and spin current $I_{s}$ (red dash line) versus the bias voltage $V$ are
shown in Fig. 3(a) at the gate voltage $V_{g}=-0.4$ mV, where the degenerate
ground states of the SMM are $\left\vert 0,2;\pm 2\right\rangle $. The spin
current $I_{s}$ is negative at most of bias voltage region except the high
positive-bias, where all the transport channels are open. The variation of
differential conductance $G$ with respect to the bias voltage $V$ is plotted
in Fig. 3(b), which displays three sequential resonant peaks in the positive
bias. The peak-$1$ corresponds to the spin-down sequential transition $%
\left\vert 0,2;-2\right\rangle \Leftrightarrow \left\vert
1,5/2;-5/2\right\rangle $, and the peak-$2$ (peak-$3$) corresponds to the
spin-up transition $\left\vert 0,2;-2\right\rangle \Leftrightarrow
\left\vert 1,3/2;-3/2\right\rangle $ ($\left\vert 1,5/2;-5/2\right\rangle
\Leftrightarrow \left\vert 2,2;-2\right\rangle $). In the negative bias, the
peak-$1^{\prime }$ corresponds to the spin-up transition $\left\vert
0,2;2\right\rangle \Leftrightarrow \left\vert 1,5/2;5/2\right\rangle $, and
the peak-$2^{\prime }$ (peak-3$^{\prime }$) corresponds to the spin-down
transition $\left\vert 0,2;2\right\rangle \Leftrightarrow \left\vert
1,3/2;3/2\right\rangle $ ($\left\vert 1,5/2;5/2\right\rangle \Leftrightarrow
\left\vert 2,2;2\right\rangle $). The height of peak-$1$ is lower than that
of the peak-$1^{\prime }$, since spin-up (spin-down) electrons are majority
(minority) in the $L$-electrode.

\begin{figure}[pth]
\begin{center}
\includegraphics[width=0.9\columnwidth]{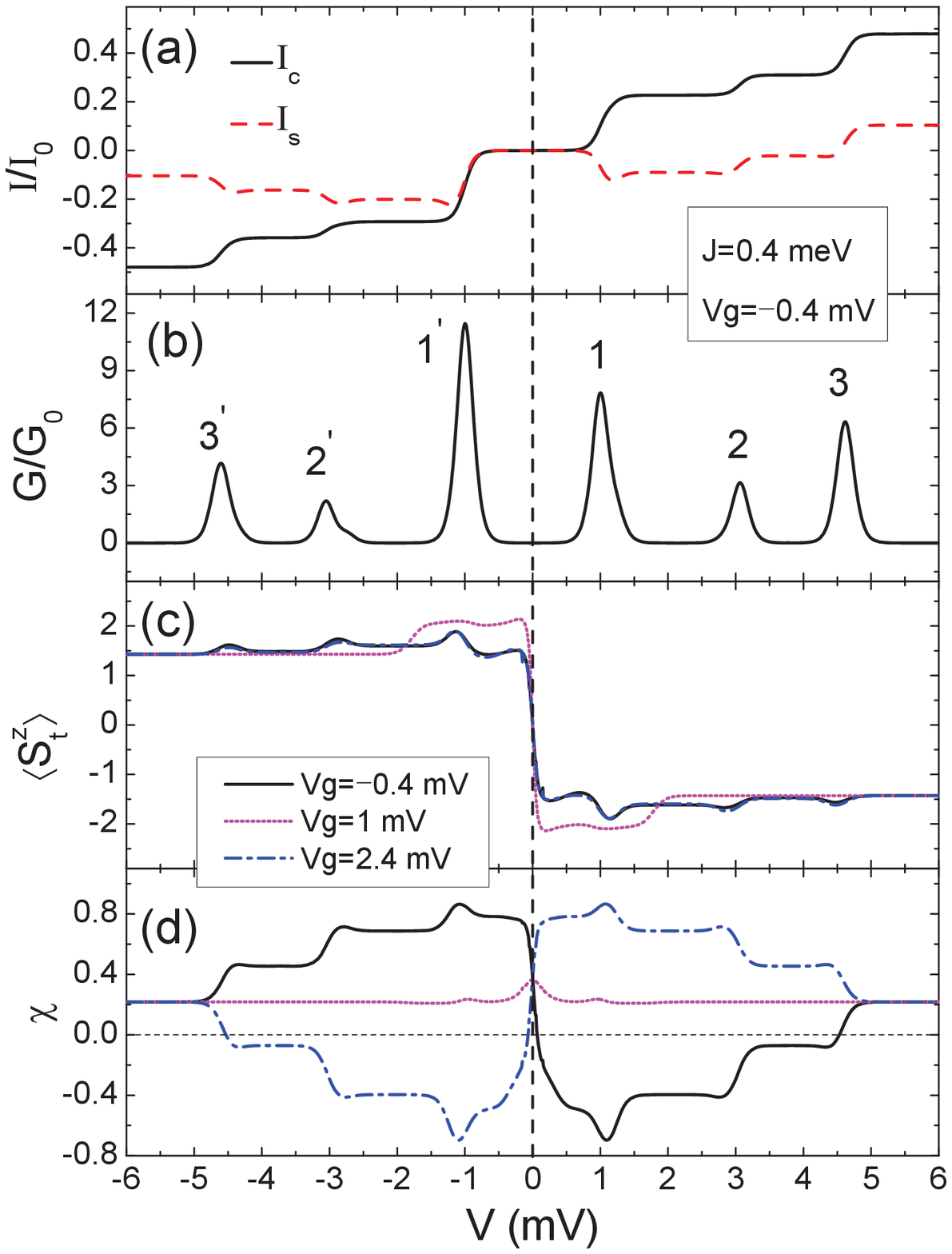}
\end{center}
\caption{(Color online) In the case of FM exchange coupling ($J=0.4$ meV):
(a) charge current $I_{c}$ and spin current $I_{s}$, (b) differential
conductance $G$, (c) magnetization $\left\langle S_{t}^{z}\right\rangle $,
and (d) spin-current polarization $\protect\chi$ as a function of the bias
voltage $V $ for different gate voltages $V_{g}$. The current and
differential conductance are scaled in units of $I_{0}=2e\Gamma /\hbar $ and
$G_{0}=10^{-3}e^{2}/h$, respectively.}
\label{Fig:3}
\end{figure}

\begin{figure}[pth]
\begin{center}
\includegraphics[width=0.9\columnwidth]{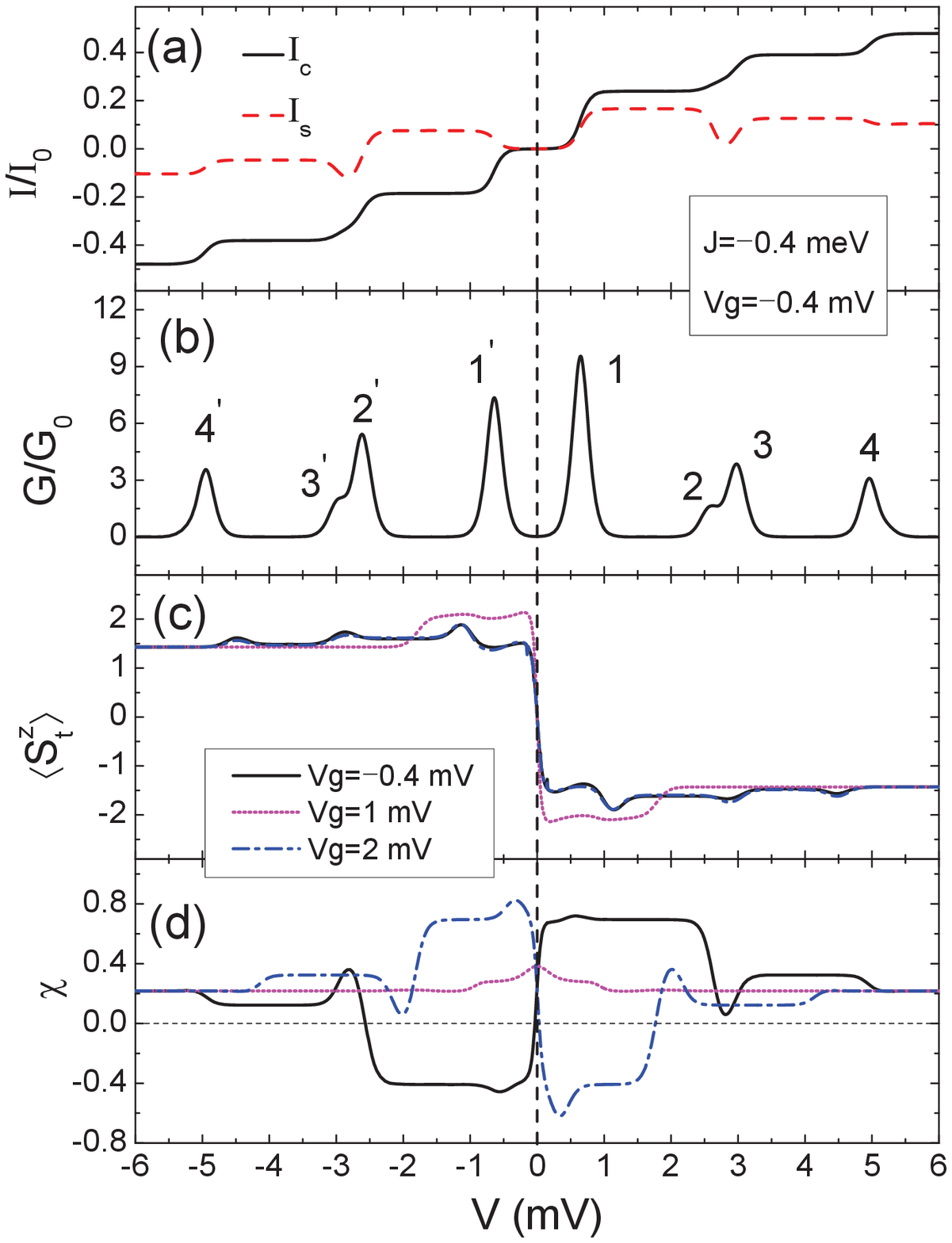}
\end{center}
\caption{(Color online) In the case of AFM exchange coupling ($J=-0.4$ meV):
(a) charge current $I_{c}$ and spin current $I_{s}$, (b) differential
conductance $G$, (c) magnetization $\left\langle S_{t}^{z}\right\rangle $,
and (d) spin-current polarization $\protect\chi$ as a function of the bias
voltage $V $ for different gate voltages $V_{g}$.}
\label{Fig:4}
\end{figure}

The spin-current polarization $\chi $ as a function of the bias voltage $V$
is shown in Fig. 3(d) for different gate voltages. At low bias voltages
around $V=0$ mV, the electron transport is dominated by elastic cotunneling
processes and thus the polarization $\chi $ approaches the spin polarization
of FM lead, namely $\chi \sim 0.4$. With the slight increase of bias voltage
$V$, the inelastic cotunneling processes start to take part in electron
transports. For the gate voltage $V_{g}=-0.4$ mV (black solid line), at
positive bias the transport current is brought mainly by the spin-down
electrons tunneling through the SMM via the virtual transition $\left\vert
0,2;-2\right\rangle \Leftrightarrow \left\vert 1,5/2;-5/2\right\rangle $,%
\cite{JBTMR1} and the polarization $\chi $ is from positive to negative
values. At negative bias, the transport is dominated by the spin-up
(spin-majority) electrons tunneling via the virtual transition $\left\vert
0,2;2\right\rangle \Leftrightarrow \left\vert 1,5/2;5/2\right\rangle $, and
the polarization $\chi $ is enhanced, larger than $0.8$. When the bias
voltage increases further to the threshold value, sequential tunneling
begins to dominate the electron transports. The polarization reaches the
lowest value $\chi \sim -0.7$ at the peak-$1$ of differential conductance as
displayed in Fig. 3 (b), where the transport is dominated by spin-down
electrons. It reduces to $-0.4$ ($-0.1$) approximately between the peak-$1$
and peak-$2$ (peak-$2$ and peak-$3$), where more spin-up electrons take part
in the transport. On the other hand, the polarization $\chi $ at the
conductance peak-$1^{\prime }$ obtains the highest value about $0.9$, where
the spin-up electrons dominate the transport. It reduces to about $0.7$ ($%
0.4 $) between the peak-$1^{\prime }$ and peak-$2^{\prime }$ (peak-$%
2^{\prime }$ and peak-$3^{\prime }$). When the absolute value of
bias-voltage $V$ is high enough, all transition channels enter the transport
window and the polarization $\chi $ remains in a constant magnitude close to
$0.2$. This situation is the same as in the QDs.\cite{QDT1} At the gate
voltage $V_{g}=1$ mV the ground states of SMM become $\left\vert 1,5/2;\pm
5/2\right\rangle $, where the spin polarization curve $\chi $ (pink dash
line) varies approximately from $0.4$ to $0.2$ with increasing bias $V$.
Since the gate voltages $V_{g}=2.4$ mV and $V_{g}=-0.4$ mV are symmetric
with respect to $V_{g}=1$ mV, which is the electron-hole symmetry point,\cite%
{JBTMR1,JBdiode} the corresponding polarization curves (blue dot-dash and
black solid lines) exhibit a symmetric behavior seen from Fig. 3(d). For $%
V_{g}=2.4$ mV, the highest polarization $\chi $ located at peak-$1$ is
mainly contributed by the spin-up transitions $\left\vert
2,2;-2\right\rangle \Leftrightarrow \left\vert 1,5/2;-5/2\right\rangle $.

Figure 4 presents the spin-polarized transport through the SMM tunnel
junction with the AFM exchange interaction between the transport electrons
and the molecular giant-spin. Different from the FM case, the polarization $%
\chi $ at the gate voltage $V_{g}=-0.4$ mV increases from $0.4$ to about $%
0.7 $ in the positive bias. This behavior is attributed to the spin-up
virtual transition $\left\vert 0,2;-2\right\rangle \Leftrightarrow
\left\vert 1,3/2;-3/2\right\rangle $. While $\chi $ decreases to the
negative value about $-0.4$ in the negative bias due to the spin-down
virtual transition $\left\vert 0,2;2\right\rangle \Leftrightarrow \left\vert
1,3/2;3/2\right\rangle $. When the bias voltage increases to a certain
higher value, the sequential tunneling dominates the electronic transport,
and the differential conductance $G$ possesses four peaks for both positive
and negative bias voltages, as shown in Fig. 4(b). The peak-$1$ mainly from
the transitions $\left\vert 0,2;-2\right\rangle \Leftrightarrow \left\vert
1,3/2;-3/2\right\rangle $ is higher than the peak-1$^{\prime }$ , which is
contributed by the transitions $\left\vert 0,2;2\right\rangle
\Leftrightarrow \left\vert 1,3/2;3/2\right\rangle $. The corresponding
polarization $\chi $ at peak-1 (peak-1$^{\prime }$) reaches the highest
(lowest) value about $0.7$ ($-0.4$). The conductance peak-$2$ and peak-$3$
are related to the transitions $\left\vert 0,2;-2\right\rangle
\Leftrightarrow \left\vert 1,5/2;-5/2\right\rangle $ and $\left\vert
1,5/2;-5/2\right\rangle \Leftrightarrow \left\vert 2,2;-2\right\rangle $
respectively, and a small dip appears for the polarization curve (black
solid line) between the two peaks. In contrast, the polarization curve is
convex between the peak-$2^{\prime }$ and peak-$3^{\prime }$. Additional
peak-$4$ and peak-$4^{\prime }$ emerge with further increase of the bias
voltage $V$. These two peaks are mainly contributed from the transitions $%
\left\vert 1,3/2;-3/2\right\rangle \Leftrightarrow \left\vert
2,2;-2\right\rangle $ and $\left\vert 1,3/2;3/2\right\rangle \Leftrightarrow
\left\vert 2,2;2\right\rangle $ respectively. More interestingly, without
the reversal of bias, the polarization $\chi $ for $V_{g}=2$ mV (blue
dot-dash line) can be reversed from about $-0.6$ to $0.4$ in positive bias
range.

\section{Conclusion}

In conclusion, we have shown that the FM-SMM-NM junction can work as an
efficient spin-current injector. By the master equation approach,
the spin-polarized transport properties are systematically investigated in
both the sequential and cotunneling regimes. Our results demonstrate that
the transport exhibits a very asymmetric behavior with respect to the zero
bias. The spin-flip process, which originates from the spin-exchange
interaction of the SMM and the transport electron, leads to the
amplification of spin polarization injecting from the FM electrode, and a
very high polarization of the spin-current is obtained. Furthermore, both
the magnitude and sign of the spin polarization are tunable by the gate or
bias voltages, suggesting an electrically-controllable spin device in
molecular spintronics.

\begin{acknowledgements}

This work was supported by National Natural Science Foundation of China
under grant Nos. 11504260, 11447163, 11574186, 11275118, 11504240, and STIP under grant Nos. 2014147, 2016170.

\end{acknowledgements}


\begin{thebibliography}{99}
\bibitem{spin} I. \v{Z}uti\'{c}, J. Fabian, and S. D. Sarma, Rev. Mod. Phys.
76, 323 (2004).

\bibitem{graphene1} W. Han, K. Pi, K. M. McCreary, Y. Li, J. J. I. Wong, A.
G. Swartz, and R. K. Kawakami, Phys. Rev. Lett. 105, 167202 (2010).

\bibitem{graphene2} C. Zhang, Y. Wang, B. Wu, and Y. Wu, Appl. Phys. Lett.
101, 022406 (2012).

\bibitem{silicon} L. K. Castelano and L. J. Sham, Appl. Phys. Lett. 96,
212107 (2010).

\bibitem{QDE1} C. A. Merchant and N. Markovic, Phys. Rev. Lett. 100, 156601
(2008).

\bibitem{QDE2} K. Hamaya, M. Kitabatake, K. Shibata, M. Jung, S. Ishida, T.
Taniyama, K. Hirakawa, Y. Arakawa, and T. Machida, Phys. Rev. Lett. 102,
236806 (2009).

\bibitem{QDT1} F. M. Souza, J. C. Egues, and A. P. Jauho, Phys. Rev. B 75,
165303 (2007).

\bibitem{QDT2} S. Csonka, I. Weymann, and G. Zarand, Nanoscale 4, 3635
(2012).

\bibitem{sp} R. Meservey and P. M. Tedrow, Phys. Rep. 238, 173 (1994).

\bibitem{Wernsdorfer} L. Bogani and W. Wernsdorfer, Nature Mater. 7, 179
(2008).

\bibitem{exp1} H. B. Heersche, Z. de Groot, J. A. Folk, H. S. J. van der
Zant, C. Romeike, M. R. Wegewijs, L. Zobbi, D. Barreca, E. Tondello, and A.
Cornia, Phys. Rev. Lett. 96, 206801 (2006).

\bibitem{Kim} C. Romeike, M. R. Wegewijs, and H. Schoeller, Phys. Rev. Lett.
96, 196805 (2006).

\bibitem{Spinfilter1} L. Zhu, K. L. Yao, and Z. L. Liu, Appl. Phys. Lett.
96, 082115 (2010).

\bibitem{Spinfilter2} H. Hao, X. H. Zheng, Z. X. Dai, and Z. Zeng, Appl.
Phys. Lett. 96, 192112 (2010).

\bibitem{Xing2} R.-Q. Wang, L. Sheng, R. Shen, B. Wang, and D.Y. Xing, Phys.
Rev. Lett. 105, 057202 (2010);

\bibitem{Xing3} Z. Zhang, L. Jiang, R. Wang, B. Wang, and D. Y. Xing, Appl.
Phys. Lett. 97, 242101 (2010).

\bibitem{JBTMR1} M. Misiorny and J. Barna\'{s}, Phys. Rev. B 79, 224420
(2009);

\bibitem{Xie1} H. Xie, Q. Wang, H. B. Xue, H. Jiao and J.-Q. Liang, J. Appl.
Phys. 113, 213708 (2013).

\bibitem{JBTMR2} M. Misiorny, I. Weymann, and J. Barna\'{s}, Phys. Rev.
Lett. 106, 126602 (2011)

\bibitem{JBTMR3} M. Misiorny, I. Weymann, and J. Barna\'{s}, Phys. Rev. B
84, 035445 (2011).

\bibitem{Timm3} C. Timm and M. Di Ventra, Phys. Rev. B 86, 104427 (2012).

\bibitem{SSeebeck} M. Misiorny and J. Barna\'{s}, Phys. Rev. B 89, 235438
(2014).

\bibitem{SSeebeck1} M. Misiorny and J. Barna\'{s}, Phys. Rev. B 91, 155426
(2015).

\bibitem{Scon} A. P{\l }omi\'{n}ska, M. Misiorny and J. Barna\'{s}, Phys.
Rev. B 95, 155446 (2017).

\bibitem{Timm1} C. Timm and F. Elste, Phys. Rev. B 73, 235304 (2006).

\bibitem{JB} M. Misiorny and J. Barna\'{s}, Phys. Rev. B 76, 054448 (2007).

\bibitem{Rossier} F. Delgado and J. Fern\'{a}ndez-Rossier, Phys. Rev. B 82,
134414 (2010).

\bibitem{Xing1} Z. Zhang, L. Jiang, R. Wang, B. Wang, and D. Y. Xing, Appl.
Phys. Lett. 99, 133110 (2011).

\bibitem{JB1} M. Misiorny and J. Barna\'{s}, Phys. Rev. Lett. 111, 046603
(2013).

\bibitem{exp2} S. Loth, K. von Bergmann, M. Ternes, A. F. Otte, C. P. Lutz,
and A. J. Heinrich, Nat. Phys. 6, 340 (2010).

\bibitem{exp3} F. D. Natterer, K. Yang, W. Paul, P. Willke, T. Choi, T.
Greber, A. J. Heinrich, and C. P. Lutz, Nature 543, 226 (2017).

\bibitem{JBdiode} M. Misiorny and J. Barna\'{s}, Europhys. Lett. 89, 18003
(2010).

\bibitem{Timm2} F. Elste and C. Timm, Phys. Rev. B 73, 235305 (2006).

\bibitem{Xue} H.-B. Xue, J.-Q. Liang, and W.-M. Liu, Sci. Rep. 5, 8730
(2015).

\bibitem{Xie2} H. Xie, Q. Wang, H. Jiao, and J.-Q. Liang, J. Appl. Phys.
112, 043701 (2012).

\bibitem{Xie3} H. Xie, Q. Wang, B. Chang, H. Jiao, and J.-Q. Liang, J. Appl.
Phys. 111, 063707 (2012).

\bibitem{Shen} H.-Z. Lu, B. Zhou, and S.-Q. Shen, Phys. Rev. B 79, 174419
(2009).

\bibitem{Timm5} C. Timm, Phys. Rev. B 77, 195416 (2008).

\bibitem{Koch2} J. Koch, F. von Oppen, and A. V. Andreev, Phys. Rev. B 74,
205438 (2006).

\bibitem{Bruus} H. Bruus and K. Flensberg, \textit{Many-body Quantum Theory
in Condensed Matter Physics} (Oxford University Press, Oxford, 2004).
\end{thebibliography}
\end{document}